\documentclass[aps,pra,twocolumn,superscriptaddress,reprint, showpacs]{revtex4-1}
\usepackage{graphicx}
\usepackage{setspace}
\usepackage{amsmath}
\usepackage{amssymb}
\usepackage{color}
\usepackage{array}
\usepackage{subfigure}
\usepackage{hyperref}
\usepackage{float}
\usepackage{lipsum}

\usepackage[all]{xy}
\newcommand{\RN}[1]{%
  \textup{\uppercase\expandafter{\romannumeral#1}}%
}

\begin{document}
%\begin{CJK*}{GB}{}
\title{Analysis of quantum interface between Rydberg-blocked atomic ensemble and cavity optical field with two-photon transition}

\author{Yuan Sun}
\email[email: ]{sunyuan17@nudt.edu.cn}
\affiliation{Interdisciplinary Center for Quantum Information, National University of Defense Technology, Changsha 410073, P.R.China}
\affiliation{Key Laboratory of Quantum Optics and Center of Cold Atom Physics, Shanghai Institute of Optics and Fine Mechanics, Chinese Academy of Sciences, Shanghai 201800, P.R.China}
\author{Ping-Xing Chen}
\affiliation{Interdisciplinary Center for Quantum Information, National University of Defense Technology, Changsha 410073, P.R.China}

\begin{abstract}
We study the atom-photon quantum interface with intracavity Rydberg-blocked atomic ensemble where the ground-Rydberg transition is realized by two-photon transition. Via theoretical analysis, we report our recent findings of the Jaynes-Cummings model on optical domain and robust atom-photon quantum gate enabled by this platform. The requirement on the implementation is mild which includes an optical cavity of moderately high finesse, typical alkali atoms such as Rb or Cs and the condition that cold atomic ensemble is well within the Rydberg blockade radius. The analysis focuses on the atomic ensemble's collective coupling to the quantized optical field in the cavity mode. We demonstrate its capability to serve as a controlled-PHASE gate between photonic qubits and matter qubits. The detrimental effects associated with several major decoherence factors of this system are also considered in the analysis.
\end{abstract}


\maketitle
%\end{CJK*}

\section{Introduction}
\label{sec:introduction}

Efficient quantum interface between atom and light is one of the central topics in the research frontier of quantum optics over the last three decades, which is not only important for the theoretical investigations of quantum electrodynamics (QED), but also crucial for applications in quantum information processing \cite{RevModPhys.74.347} and quantum metrology \cite{nphoton.2011.35}. In particular, two major types of experimental platforms have attracted a lot of intense attention and turn out to be spectacularly fruitful, which are the cavity quantum electrodynamics (cQED) approach and the atomic ensemble approach. The cQED approach is primarily within the framework of one or a few isolated neutral atoms coupled to a high finesse optical cavity \cite{RevModPhys.87.1379}, where the second quantization of the intracavity optical field is essential. The atomic ensemble approach \cite{RevModPhys.82.1041} is a powerful alternative to the cQED approach, where many atoms are coerced to form a collective superposition state to enhance the atom-photon coupling.

Many exciting progresses have been achieved so far, including the advent of atom-photon controlled-PHASE gate \cite{PhysRevLett.92.127902, Reiserer13177} and the photon-photon quantum gate \cite{nature18592}, the quantum nondemolition measurement of single photon pulses \cite{Reiserer1349}, the quantum memory for light \cite{nature03064, Choi06670}, the quantum repeater for photonic polarization entanglement \cite{RevModPhys.83.33}, and the quantum networking between matter qubits \cite{Kimble07127, Ritter12, PhysRevA.89.022317}, to name a few. Among many potential technological breakthroughs, the phenomenon of Rydberg blockade \cite{PhysRevLett.87.037901} is widely perceived to be quite promising in enhancing the strength of single-photon's coupling to atoms. 

The rapid progress in the research of Rydberg-Rydberg interactions of the past two decades \cite{RevModPhys.82.2313}, both theoretical and experimental, has already found many key advances in quantum optics with neutral atoms and single-photon pulses \cite{PhysRevA.66.065403, PhysRevLett.99.260501, PhysRevLett.107.093601, PhysRevLett.107.133602, PhysRevLett.109.233602, Dudin887Science, PhysRevLett.110.103001, PhysRevLett.113.053601, PhysRevA.92.022336, PhysRevLett.117.223001, PhysRevLett.119.160502}, where one typical feature is that many atoms' collective coupling to the optical field is amplified due to the Rydberg blockade effect. The prominent characteristics of such interactions include strong strength over a long range and coherent on-off switching, which makes it advantageous to utilize the Rydberg blockade effect with atomic ensembles \cite{Dudin887Science, PhysRevLett.115.093601, J.Phys.B.49.202001}.

Especially, Rydberg blockade effect is known to be a powerful tool in the study of the Jaynes-Cummings model (JCM) \cite{JCM1443594}, where the field part of JCM is either instantiated by cavity optical modes \cite{PhysRevA.82.053832} or via isomorphism to collective atomic excitations \cite{PhysRevA.66.032109, PhysRevLett.100.170504, PhysRevA.86.023845, PhysRevA.90.043413, PhysRevLett.117.213601}. JCM establishes the general framework for the interaction between a single two-level atom and quantized electromagnetic field, and it is at the very heart of the cQED platform. The system of intracavity Rydberg-blocked atomic ensemble provides an interesting opportunity to combine both the cQED and atomic ensemble techniques for the atom-light quantum interface \cite{PhysRevA.82.053832}. Optical nonlinearities have already been experimentally observed in cold atomic ensemble with strong Rydberg-Rydberg interactions inside moderate finesse cavity \cite{PhysRevLett.109.233602}, which is proven to be a natural consequence of Rydberg blockade shift \cite{RevModPhys.82.2313, PhysRevA.92.043841}. These efforts have demonstrated the unique capability of intracavity Rydberg-blocked atomic ensemble, and paved the way for further development. Thanks to the technical progress of manipulating cavity Rydberg polaritons \cite{PhysRevA.93.041802}, recent experimental efforts have confirmed JCM's specific signatures in such systems \cite{PhysRevA.95.041801, PhysRevLett.109.233602, 1367-2630-16-4-043020}.

Nevertheless, these theoretical investigations and experimental demonstrations so far, while being physically isomorphic to the JCM, have mainly focused on the extraction of quantum optical characteristics from the system's response to classical light input \cite{PhysRevA.82.053832, PhysRevA.90.043413}. Relatively much less attention has been devoted to the discussion of explicit and direct operations of such a system with a realistically quantized optical field where the JCM's input-output coupling is associated with single-photon or few-photon pulses. On the other hand, there is now a growing interest in constructing novel quantum optical devices with single-photon level manipulation capabilities via Rydberg-Rydberg interactions \cite{PhysRevLett.107.093601, PhysRevX.7.031007, PhysRevLett.120.113601}, which strongly demands further explorations of JCM with Rydberg-blocked ensemble under the condition of truly quantized optical field. Meanwhile, recent scientific advances \cite{PhysRevLett.113.133601} have confirmed the potential technical ability to precisely manipulate the genuine quantized intracavity optical state. All these considerations naturally lead to a push for a further study which would straightforwardly reveal the quantum mechanical nature of the JCM on the optical domain with intracavity Rydberg-blocked atomic ensemble. The dynamics of such a hybrid system involves collective excitations of an atomic ensemble, which is inherently related to Dicke superradiance \cite{PhysRev.93.99, PhysRevLett.102.143601}, an intriguing effects in quantum optics. Moreover, the effort along this line is capable of exploring the physical process of cQED within a parameter space not easily accessible via coupling a single atom to a high-finesse optical cavity.

Designing robust atom-photon controlled-PHASE gate is a primary mission in the research of atom-light quantum interface. It is widely known that the intracavity JCM on optical domain is very closely tied to such a quantum gate \cite{PhysRevLett.92.127902}, and therefore the study of JCM with intracavity Rydberg-blocked atomic ensemble is naturally expected to be relevant as well. Indeed, recent investigations have revealed that novel controlled-PHASE gates for atom-photon, atom-atom, photon-photon systems can be built via the help of Rydberg mediated interactions and intracavity Rydberg EIT \cite{PhysRevLett.112.040501, PhysRevA.91.030301, PhysRevLett.119.113601}. In particular, such ideas of constructing controlled-Z (C-Z) gate has ingeniously improved the effective atom-photon coupling strength by utilizing Rydberg blockade among many atoms \cite{Hao2015srep, PhysRevA.93.040303, PhysRevA.94.053830}. These new proposals are conveniently compatible with the current mainstream experimental techniques of trapping cold atom ensemble inside an optical cavity while greatly reduces the requirement on the cavity finesse as the core advantage.

In this article, we report our latest findings for the system of intracavity Rydberg-blocked atomic ensemble on its relations with the JCM and the atom-photon quantum gate. We analyze in theory the two-photon interactions between Rydberg-blocked atomic ensemble and quantized intracavity optical fields with collective coupling, where the two-photon transition is composed of the intracavity photon and one external control laser. We study the JCM with the preloaded intracavity single-photon and few-photon states as the initial condition, enabled by the special characteristics of this system. Thereafter an atom-photon controlled-PHASE quantum gate is proposed, deriving from the dressed property and Rydberg blockade effect on top of the JCM dynamics embedded in this system. One prominent feature of this gate design is that it allows for a considerable amount of flexibility on the frequency of the incident single-photon pulse.

The rest of the article is organized according to the following structure. In Section \ref{sec:framework}, we offer an overall sketch of the involved physical process. In Section \ref{sec:JCM}, the time evolution of JCM is discussed, where the coupling between the intracavity field and the outside field is taken into consideration. In Section \ref{sec:gate}, the atom-photon quantum gate is discussed. Section \ref{sec:conclusion} concludes the article.

\section{Overview and fundamentals}
\label{sec:framework}

\begin{figure}[htbp]
\centering
\includegraphics[width=\linewidth]{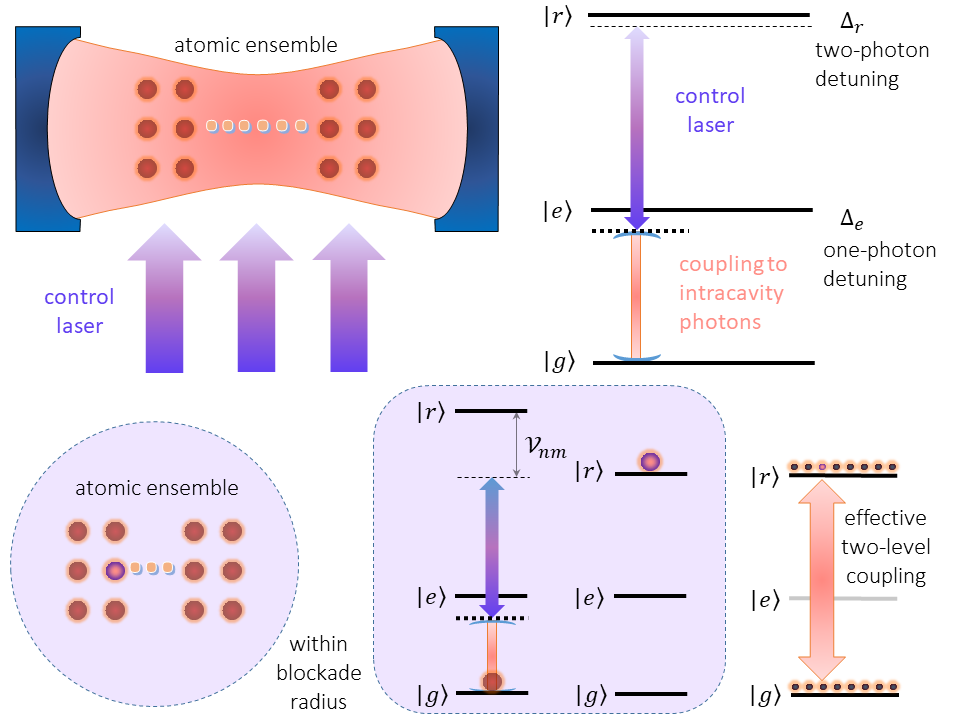}
\caption{Schematic of the intracavity Rydberg-blocked atomic ensemble system under investigation. The size of the entire ensemble spans on the order of a few tens of micrometers, which is compatible with the experimentally attainable Rydberg blockade radius. Moreover, in such a configuration the atomic ensemble well matches the cavity mode spatially, which conveniently serves for the purpose of atom-photon coupling. The control laser is of frequency $\omega_c$, while the cavity resonance frequency is $\omega_d$.}
\label{fig:basic_0}
\end{figure}

In this section, we sketch the basic physical system under investigation and present the rudimentary ingredients of the theoretical analysis that will follow. To begin with, the general setting of the intracavity Rydberg-blocked atomic ensemble is outlined in Fig. \ref{fig:basic_0}. The atomic ensemble is supposed to contain a few hundred up to a few thousand atoms, preferably in the array configuration \cite{PhysRevA.79.012320, PhysRevA.81.043822} via far off-resonance trapping (FORT) in optical lattice, such as the well-established experimental platforms of a 2D array \cite{PhysRevA.92.022336} or 3D array \cite{Weiss2581Science}. It is necessary for the geometric configuration to ensure that the entire ensemble fits within the Rydberg blockade radius for the target Rydberg states, which can be achieved by the trapping site spacing on the oder of half a micrometer \cite{PhysRevA.79.012320} for the atomic array configuration. Nevertheless, as long as the atoms' temperature is cold enough, it is not necessary to prepare the atomic ensemble in the array configuration since single atom addressing is not required for the purpose here \cite{PhysRevA.82.053832, PhysRevA.93.040303}.

For this section, the cavity mirrors are assumed to have 100\% perfect reflectivity and therefore the intracavity optical field won't leak to the outside. The interaction Hamiltonian to describe such a hybrid system, including the effect of Rydberg blockade in the form of van der Waals interaction potential \cite{RevModPhys.82.2313, PhysRevA.93.040303, PhysRevA.92.042710}, is given below. 
\begin{align}
\label{H_vdW_0}
H_\text{int} = 
&\sum_{n=1}^{N}(-\frac{\hbar\Omega}{2}|r_n\rangle\langle e_n| - i\hbar \mathcal{G}_n|e_n\rangle\langle g_n| \hat{b}) + \text{H.c.} \nonumber\\ % couplings
&-\hbar\Delta_e\sum_{n=1}^{N} |e_n\rangle\langle e_n| 
-\hbar\Delta_r\sum_{n=1}^{N} |r_n\rangle\langle r_n| \nonumber\\ % effective detunings
&+ \hbar\sum_{n=1}^{N}\sum_{m>n} \mathcal{V}_{nm}|r_n\rangle\langle r_n| \otimes |r_m\rangle\langle r_m|, % Rydberg blockade
\end{align}
where the one-photon detuning is $\Delta_e = \omega_c - (\omega_e - \omega_g)$ and the two-photon detuning is $\Delta_r = \omega_c + \omega_d - (\omega_r - \omega_g)$.

In \eqref{H_vdW_0}, the Rydberg-Rydberg interaction is effectively described by the potential $\mathcal{V}_{nm}$ where the details of dipole-dipole interaction is hidden for the sake of simplicity. We discuss the dynamics of the system under the condition of strong Rydberg blockade such that the doubly occupied Rydberg state like $|r_n r_m\rangle$ is never populated. Within the scope of our study where the ground state is ultimately linked to the Rydberg state with adequate strength around resonance point, the implicit assumption is that at most only a single excitation into the Rydberg state is practically allowed, which is a widely applicable approximation if $|\mathcal{V}_{nm}|$ is sufficiently large. Therefore the Hamiltonian can be reduced to:
\begin{align}
\label{H_0}
H_\text{int} = 
& (-\frac{\hbar\Omega}{2}|\tilde{r}\rangle\langle \tilde{e}| - i\hbar \mathcal{G} |\tilde{e}\rangle\langle g^N| \hat{b}) + \text{H.c.} 
\nonumber\\ % couplings
&-\hbar\Delta_e |\tilde{e}\rangle\langle \tilde{e}| 
-\hbar\Delta_r |\tilde{r}\rangle\langle \tilde{r}|,  % effective detunings
\end{align}
with the single excitation state defined as:
\begin{equation}
\label{superradiance_basis_0}
|\tilde{e}\rangle = \mathcal{G}^{-1}\sum \mathcal{G}_n |e_n\rangle,\,
|\tilde{r}\rangle = \mathcal{G}^{-1}\sum \mathcal{G}_n |r_n\rangle;
\end{equation}
where the normalization factor is the collective atom-photon coupling strength $\mathcal{G} = (\sum |\mathcal{G}_n|^2)^{1/2}$ and $|g^N\rangle$ stands for all N atoms in the ground state $|g\rangle$.

For the situation of $|\Delta_e|$ much larger than $\mathcal{G}$ and $|\Omega|$ while the optical field contain at most a handful of photons, upon considerations on the operator side we can carry out the adiabatic elimination under the approximation that $-\frac{\hbar\Omega}{2}|\tilde{r}\rangle\langle \tilde{e}| + i\hbar \mathcal{G} |g^N\rangle\langle \tilde{e}| - \hbar\Delta_e |\tilde{e}\rangle\langle \tilde{e}| \approx 0$. this leads to the simple linear relation:
$|\tilde{e}\rangle \approx (\hbar\Delta_e)^{-1}(-\frac{\hbar\Omega}{2}|\tilde{r}\rangle + i\hbar \mathcal{G} |g^N\rangle)$. The details of the adiabatic elimination procedure is a well established subject in the literature (also see \href{https://www.osapublishing.org/}{Supplement 1}). Consequently, the isomorphism to JCM is then readily obtained via combining this relation with \eqref{H_0}, which leads to the effective two-level system Hamiltonian:
\begin{align}
\label{H_eff_1}
H_\text{eff} = 
& \hbar \frac{\Omega\mathcal{G}}{2\Delta_e}|\tilde{r}\rangle\langle g^N| + \text{H.c.}  \nonumber\\ % couplings
&+ \frac{\mathcal{G}^2}{\Delta_e}|g^N\rangle\langle g^N| \hat{b}^\dagger\hat{b}
-\hbar(\Delta_r - \frac{\Omega^2}{4\Delta_e}) |\tilde{r}\rangle\langle \tilde{r}|. % effective detunings
\end{align}
where it is identical to the Hamiltonian for JCM of single-atom coupling to optical fields with coupling strength $|\frac{\mathcal{G} \Omega^*}{2\Delta_e}|$, apart from the ac Stark shift terms of $-i\frac{\mathcal{G}^2}{\Delta_e}C_b(t)$ and $-i\frac{|\Omega|^2}{4\Delta_e}$. Another feature is that the exact number of atoms does not matter in this type of instantiation of JCM.

This elementary model analysis is relatively straightforward, while it sketches the theme of the JCM for such a hybrid system where the single emitter is replaced by many emitters interacting collectively, thanks to superradiance and Rydberg blockade. The dynamics is certainly more complicated when the cavity is coupled to the outside optical fields, especially if the focus is put on the quantum nature of the system's response. And that is the direction we are heading for in the next two sections, including the revisit of JCM with initial condition of intracavity quantum optical fields and an efficient atom-photon quantum gate.

\section{JCM on optical domain}
\label{sec:JCM}

\subsection{Dynamics with single-photon state}
\label{sec:basic_model}

\begin{figure}[h]
\centering
\includegraphics[width=\linewidth]{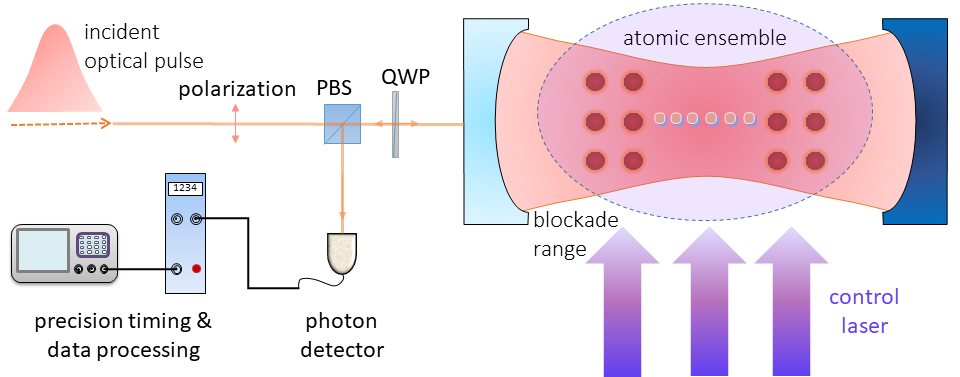}
\caption{Schematic of the system for the study of JCM with intracavity Rydberg blocked atomic ensemble, where the state initialization includes the process of feeding non-classical optical pulse into the cavity. The optical cavity is supposed to be single-sided, where the one end mirror is perfectly reflecting. In this particular example, the feeding and retrieving of the intracavity optical field is realized by polarization optics for the free space optical pulse. Typical parameter settings for the experimental implementation is relatively mild for the hardware nowadays. For example, the cavity finesse can be set as $\mathcal{F}\approx5\times10^3$, the cavity decay rate can be chosen as $\kappa \approx 2\pi \times 0.5$ MHz, the cavity free spectral range can be set as $\textit{FSR} \approx 5$ GHz, and the number of atoms in the ensemble can be chosen as $\sim$100--1000. The instantiation of the Jaynes-Cummings model in such a system is relatively straightforward compared with the case of high finesse cavity, thanks to the recent developments in the single photon pulse engineering and Rydberg atom control techniques.}
\label{fig:basic_JCM}
\end{figure}

Here we begin to treat the JCM dynamics with input-output coupling to the free space modes for the intracavity Rydberg-blocked atomic ensemble system sketched in Section \ref{sec:framework}. An example of the implementation is illustrated in Fig. \ref{fig:basic_JCM}. The situation under study is that a single-photon state for the intracavity field is prepared as the initial condition, and then the JCM dynamics mandated by \eqref{H_vdW_0} and \eqref{H_eff_1} is consequently invoked. 

The initial condition can be realized by efficiently and deterministically loading a prescribed single-photon optical pulse into a cavity \cite{PhysRevLett.113.133601}. This loading process is supposed to be carried out with the control laser shut off such that the intracavity optical field is off-resonant with the atomic transition $|g\rangle \to |e\rangle$. That is to say, the cavity seems to be empty for the incoming optical field when $\Delta_e \gg \mathcal{G}$. As soon as the deterministic single-photon loading stage is completed, the control laser turns on to enable the JCM process, which is what we are interested in.

Now that the state vector for the entire system needs to include the free space optical field component, and it reads:
\begin{align}
\label{JCM_state_vector_1}
|\Psi(t)\rangle =& \int d\omega \, \phi_s (\omega, t) \hat{a}^\dagger_s(\omega) |g^N, \O_b, \text{\O}_a\rangle  %  s stands for free Space
+ C_b(t) \hat{b}^\dagger |g^N, \text{\O}_b, \text{\O}_a\rangle
\nonumber\\
&+ \sum_{m=1}^{N} C_{em}(t) |g^{N-1}e_m, \text{\O}_b, \text{\O}_a\rangle \nonumber\\
&+ \sum_{m=1}^{N} C_{rm}(t) |g^{N-1}r_m, \text{\O}_b, \text{\O}_a\rangle.
\end{align}

Under such circumstances, the excitation of the atomic ensemble caused by the intracavity single-photon optical state is in essence a collective excitation, which can be readily observed from \eqref{JCM_state_vector_1}. The effect of superradiance is a fundamental building block for the atom-photon interaction dynamics here. Therefore, in order to examine the process in a little more detailed manner, we'd like to define the coefficients with respect to collective state basis:
\begin{equation}
\label{superradiance_basis_1}
C_e = \mathcal{G}^{-1}\sum \mathcal{G}^*_n C_{en},\,
C_r = \mathcal{G}^{-1}\sum \mathcal{G}^*_n C_{rn};
\end{equation}
where the subscript $n$ denotes the numbering of atoms.

The coupling through the cavity mirror between the intracavity mode and the free space mode can be described by the interaction Hamiltonian:
\begin{equation}
\label{input_output_H_int}
H_\text{int} = i\hbar
\int d\omega\, g_s(\omega) (\hat{a}^\dagger_s(\omega) \hat{b}
- \hat{b}^\dagger \hat{a}_s(\omega) ).
\end{equation}

According to the quantum input-output formalism (see \href{https://www.osapublishing.org/}{Supplement 1} for more details), the equations describing the dynamics can be obtained from \eqref{H_vdW_0} and \eqref{input_output_H_int}. Under the collective state basis specified in \eqref{superradiance_basis_1}, the equations of motion are formulated as:
\begin{align}
\label{JCM_EOM_1b}
\frac{d}{dt} C_b(t) &=  \mathcal{G} C_e - \frac{\kappa}{2}C_b(t),
\nonumber\\
\frac{d}{dt} C_e(t) &= -\mathcal{G} C_b(t) + i\frac{\Omega^*}{2}C_r(t) + i\Delta_e C_e(t) -\frac{\Gamma_e}{2}C_e(t),
\nonumber\\
\frac{d}{dt} C_r(t) &= i\frac{\Omega}{2}C_e(t) + i\Delta_r C_r(t) - \frac{\Gamma_r}{2}C_r(t),
\end{align}
where $\kappa$ is the effective cavity optical field decay rate, $\Gamma_e$ is the spontaneous emission rate of state $|e\rangle$, and $\Gamma_r$ is the spontaneous emission rate of state $|r\rangle$.

\begin{figure}[t!]
\centering
\includegraphics[width=\linewidth]{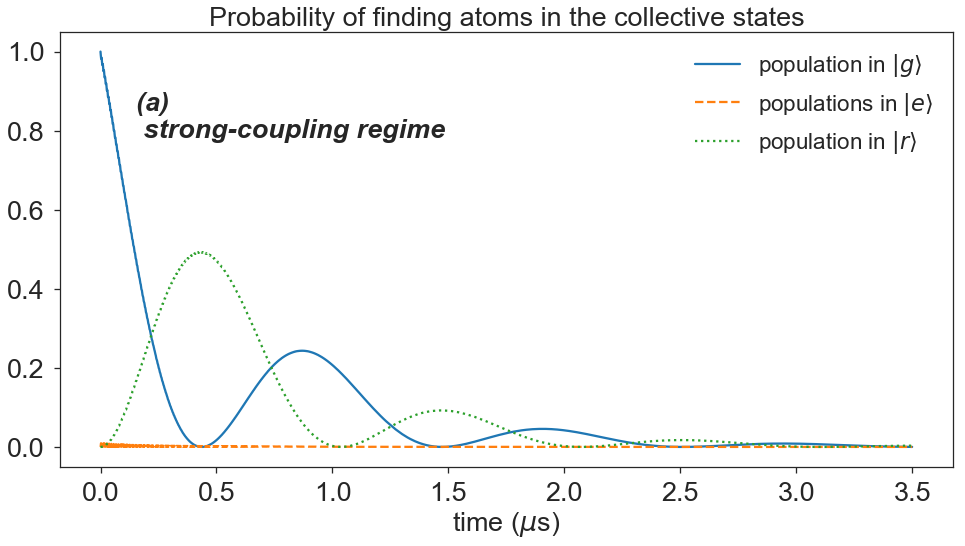}\\
\includegraphics[width=\linewidth]{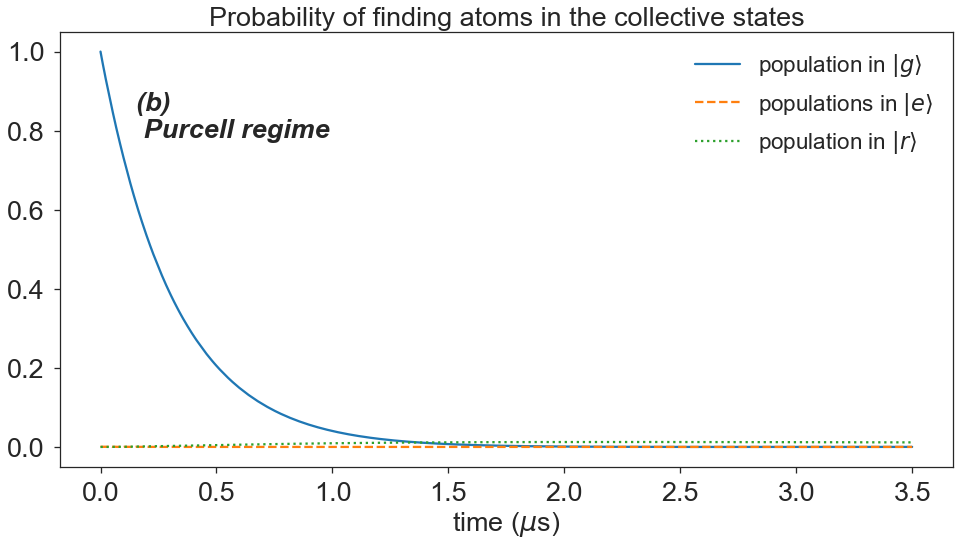}
\caption{Numerical simulation of the quantum Rabi oscillations of the intracavity atom-photon interaction, single-photon state case. Parameters are set as: $\Delta_e=2\pi \times 200 \text{ MHz}, \Delta_r = 0, \Gamma_e = 2\pi \times 1 \text{ MHz}, \Gamma_r = 2\pi \times 0.01 \text{ MHz}, \kappa = 2\pi \times 0.5 \text{ MHz}$. For (a), $\Omega = 2\pi\times 20 \text{ MHz}, \mathcal{G}=2\pi\times 10 \text{ MHz}$; for (b), $\Omega = 2\pi\times 5 \text{ MHz}, \mathcal{G}=2\pi\times 2.5 \text{ MHz}$.}
\label{fig:sample1}
\end{figure}

A numerical example is presented in Fig. \ref{fig:sample1}, according to the dynamics governed by \eqref{JCM_EOM_1b}. The signature of quantum Rabi oscillation is clear in Fig. \ref{fig:sample1}(a), as it belongs to the strong-coupling regime in the cQED terminology. The decay of the population within the system is dominated by the photon leaking out from the cavity since the Rydberg state has relatively long lifetime. In the effective two-level atom picture, this amounts to the case where the coupling to the excited state is strong while the spontaneous emission rate of the excited state is relatively much smaller. Such type of parameter setting and the initial state preparation is not easily accessible in a typical platform of a single atom coupled to high finesse cavity \cite{PhysRevLett.101.223601}.

Under the condition of a relatively large $\Delta_e$, the adiabatic elimination is again applicable which can lead to a straightforward effective two-level atom description. Analogous to \eqref{H_eff_1}, for the intermediate state we have $C_e(t)=(i\Delta_e - \frac{\Gamma_e}{2})^{-1} (\mathcal{G}C_b(t) - i\frac{\Omega^*}{2}C_r(t))$, which leads to the following equations from \eqref{JCM_EOM_1b}:
\begin{align}
\label{JCM_EOM_1c}
\frac{d}{dt} C_b(t) &= -i\frac{\mathcal{G}^2}{\Delta_e+i\frac{\Gamma_e}{2}}C_b(t) - \frac{\mathcal{G} \Omega^*}{2(\Delta_e+i\frac{\Gamma_e}{2})} C_r(t) - \frac{\kappa}{2}C_b(t),
\nonumber\\
\frac{d}{dt} C_r(t) &= \frac{\mathcal{G} \Omega}{2(\Delta_e+i\frac{\Gamma_e}{2})} C_b(t) - i\frac{|\Omega|^2}{4(\Delta_e+i\frac{\Gamma_e}{2})} C_r(t) 
\nonumber\\
&+ i\Delta_r C_r(t).
\end{align}
Apart from the ac Stark shift terms, it can be identified as the dynamics for a typical cQED system with the effective atom-photon coupling $\frac{\mathcal{G} \Omega^*}{2(\Delta_e+i\frac{\Gamma_e}{2})}$ and the cavity decay rate $\kappa$.

\subsection{Dynamics with optical coherent state}
\label{sec:coherent_states}

\begin{figure}[ht]
\centering
\fbox{\includegraphics[width=\linewidth]{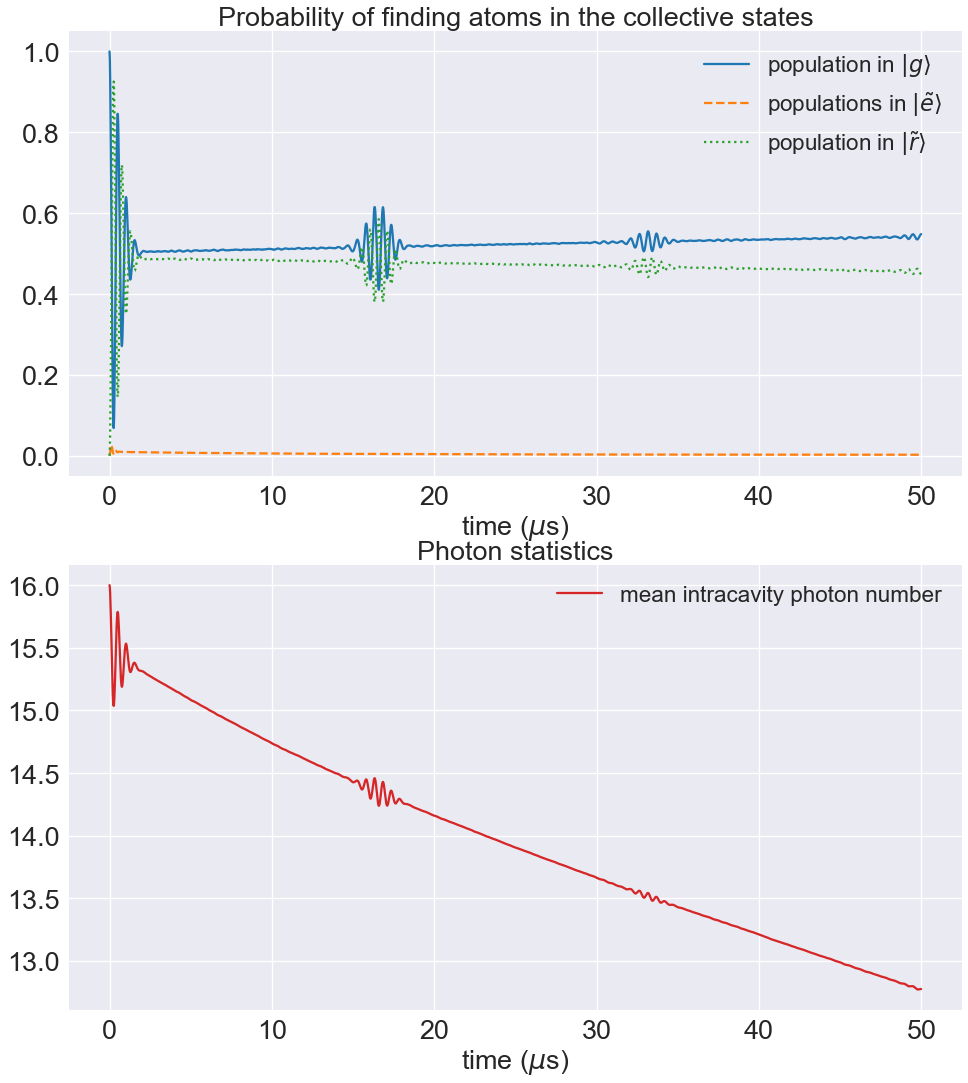}}
\caption{Numerical simulation of the quantum Rabi oscillations of the intracavity atom-photon interaction, coherent state version. The artificial condition of almost no cavity decay is imposed to show the quantum revival, while the atomic decays are retained in the calculations. Parameters are set as: $\mathcal{G}=2\pi\times2.5\text{ MHz}, \Omega=2\pi\times20\text{ MHz}, \Delta_e=2\pi \times 100 \text{ MHz}, \Delta_r = 0, \Gamma_e = 2\pi \times 1 \text{ MHz}, \Gamma_r = 2\pi \times 0.01 \text{ MHz}, \kappa = 2\pi \times 10^{-4} \text{ MHz}$. The initial condition is $|\alpha|=4$ for the optical coherent state $|\alpha\rangle$ while the cutoff is set at 50. Averaged over 10000 MCWF traces.}
\label{fig:JCM_quantum_revival}
\end{figure}

In Sub-Section \ref{sec:basic_model}, the discussion is devoted to the case where the initial condition is prepared as a deterministic single-photon state for the intracavity optical field. On the other hand, the situation of optical coherent state with small mean photon number is also frequently encountered in the study of quantum optics. It is worthwhile to investigate such a situation, with merits from the experimental side as well as the theoretical side. Practically, such an initial state is commonly prepared via a weak coherent optical pulse incident upon a cavity in configuration like Fig. \ref{fig:basic_JCM}.

The assumption of strong Rydberg blockade is kept such that at most a single excitation is allowed into the Rydberg state, throughout the rest of this section. For the contents in Sub-Section \ref{sec:basic_model}, this assumption serves pragmatically redundant since the optical field does not contain multi-photon component and therefore naturally the Rydberg excitation number can not exceed 1.

We use the quantum input-output theory together with quantum jump approach \cite{PhysRevLett.68.580, RevModPhys.70.101} to treat the time evolution of this system, which is also known as the approach of Monte-Carlo wave function (MCWF). To begin with, we extend the definition of the system's wave function coefficients $C_e, C_r$ specified in \eqref{superradiance_basis_1} to include the multi-photon case so that we set $C_{b, n}, C_{e, n}, C_{r, n}$ to denote the coefficients with respect to the Fock state basis $|n\rangle$ on the photonic side. Then the time evolution without considering any decay is given by the equations below for $n \geq 1$:
\begin{subequations}
\label{eom_JCM_multiphoton}
\begin{align}
&\frac{d}{dt} C_{b, n}(t) = \sqrt{n}\mathcal{G} C_{e, n-1}(t),
\\
&\frac{d}{dt} C_{e, n-1}(t) = -\sqrt{n}\mathcal{G} C_{b, n}(t) + i\frac{\Omega^*}{2}C_{r, n-1}(t) + i\Delta_e C_{e, n-1}(t),
\\
&\frac{d}{dt} C_{r, n-1}(t) = i\frac{\Omega}{2}C_{e, n-1}(t) + i\Delta_r C_{r, n-1}(t);
\end{align}
\end{subequations}
where $C_{b, n}, C_{e, n-1}$ and $C_{r, n-1}$ form a closed loop. In the numerical instantiation, we manually set a cutoff in $n$ depending on $|\alpha|$ from the initial condition's optical coherent state $|\alpha\rangle$. 

In particular, on top of the dynamics governed by \eqref{H_eff_1} and \eqref{eom_JCM_multiphoton}, MCWF is going to also take the decays into consideration in order to numerically resolve the time evolution. 
\emph{Atomic spontaneous emissions $\Gamma_r, \Gamma_e$}. For a small time interval $\Delta t$, the probability of detecting spontaneous emission from $|e\rangle$ is $\Gamma_e \times \Delta t \times \sum_{n=0}^{N}|C_{e, n}|^2$, while the chance of detecting $\omega_r$ is $\Gamma_r \times \Delta t \times \sum_{n=0}^{N}|C_{r, n}|^2$. If quantum jumps from the atomic spontaneous emissions do take place, then the atomic state is dragged into the specific eigenstates, namely the ground state. Note that a spontaneously emitted single-photon pulse of $\omega_e$ or $\omega_r$ won't yield information about the wave function of the optical field part, therefore it is rightful to expect the wave function of the optical field part remains untouched for the quantum jump. For example, if the quantum jump of $\Gamma_e$ does occur in one trial of MCWF, then the wave function of the system shall become a series of only $[C_{b, 0}, C_{b, 1}, C_{b, 2}, \cdots, C_{b, n}, \cdots, C_{b, N-1}]$ with values from $[C_{e, 0}, C_{e, 1}, C_{e, 2}, \cdots, C_{e, n}, \cdots, C_{e, N-1}]$ prior to the quantum jump. The situation of quantum jump associated with $\Gamma_r$ is analogous. If quantum jumps do not take place, the corresponding excited-state amplitudes shall be reduced. That is to say, $1-\frac{\Gamma_e}{2}\Delta t$ for $C_{e, n}$'s and $1-\frac{\Gamma_r}{2}\Delta t$ for $C_{r,n}$'s. 
\emph{Cavity optical field decay $\kappa$}. For a small time interval $\Delta t$, the chance of detecting a single-photon pulse from cavity emission is $\kappa \times \Delta t \times \sum_{n=1}^{N}n(|C_{b,n}|^2+|C_{e,n}|^2 + |C_{r,n}|^2)$, where the factor $n$ is from the property of the annihilation operator $\hat{a}|n\rangle = \sqrt{n}|n-1\rangle, n>1$. If quantum jumps do take place, note that by detecting a single-photon pulse from the cavity emission won't yield information on which Fock state the cavity optical field is sitting at. Therefore, conditioned on such a quantum jump, the change to the wave function of the entire atom-photon hybrid system can be understood as:
$C_{b,n} \to \sqrt{n}C_{b,n-1}, C_{e,n} \to \sqrt{n}C_{e,n-1}, C_{r,n} \to \sqrt{n}C_{r,n-1}$ for $n \geq 1$. On the other hand, if quantum jumps do not take place, then the amplitude of those states with non-zero intracavity photon components shall be reduced accordingly:
$C_{b,n} \to (1-\frac{\kappa n}{2}\Delta t)C_{b,n}, C_{e,n} \to (1-\frac{\kappa n}{2}\Delta t)C_{e,n}, C_{r,n} \to (1-\frac{\kappa n}{2}\Delta t)C_{r,n}$ for $n \geq 1$.

\begin{figure}[ht]
\centering
\fbox{\includegraphics[width=\linewidth]{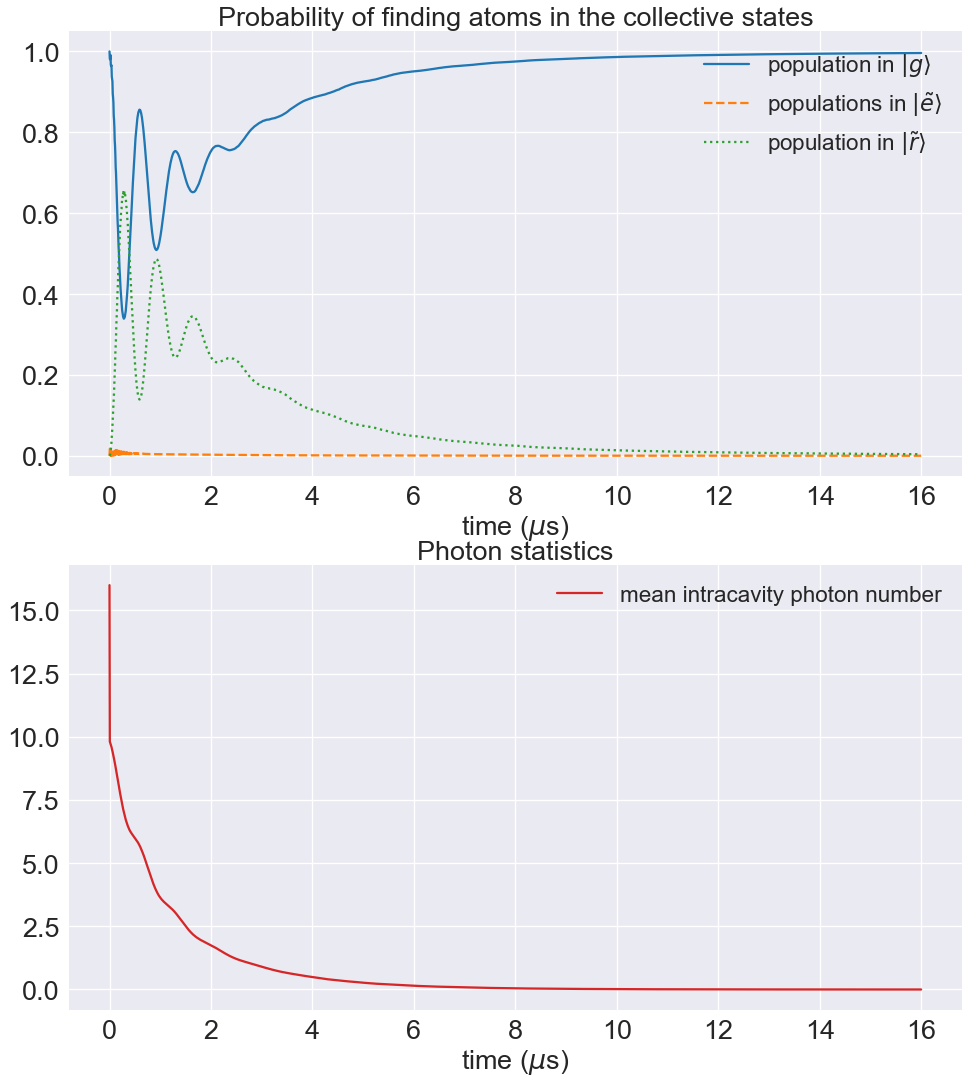}}
\caption{Numerical simulation of the quantum Rabi oscillations of the intracavity atom-photon interaction, coherent state version. A  moderate cavity decay into the free space environment is considered here. Parameters are set as: $\mathcal{G}=2\pi\times2.5\text{ MHz}, \Omega=2\pi\times20\text{ MHz}, \Delta_e=2\pi \times 100 \text{ MHz}, \Delta_r = 0, \Gamma_e = 2\pi \times 1 \text{ MHz}, \Gamma_r = 2\pi \times 0.01 \text{ MHz}, \kappa = 2\pi \times 0.1 \text{ MHz}$. The initial condition is $|\alpha|=4$ for the optical coherent state $|\alpha\rangle$ while the cutoff is set at 50. Averaged over 10000 MCWF traces.}
\label{fig:JCM_typical_decay}
\end{figure}

With the above prescription, together with the final stage of re-normalization of the wave function at the end of each time interval $\Delta t$, MCWF is able to numerically yield the result if the initial condition is specified. 
If the initial intracavity optical field is prepared as a coherent state with mean photon number not too large, the phenomenon of collapse and revival of quantum Rabi oscillation is anticipated when the optical cavity decay is very small. An example of numerical simulation for such case is shown in Fig. \ref{fig:JCM_quantum_revival}. However, from a practical point of view, such small $\kappa$ is hard to achieve experimentally.
For a reasonable finite value of $\kappa$, the dynamics is different. Typically, after some short period of time, the intracavity photon decreases by a lot due to the loss from cavity emission.  This changes the amount of ac Stark shift induced by the intracavity optical field, and ultimately causes the two-photon transition to the Rydberg state to be off-resonant. See the subsequent numerical simulation result in Fig. \ref{fig:JCM_typical_decay} for an example.
The common feature of the relatively fast oscillations in both Fig. \ref{fig:JCM_quantum_revival} and Fig. \ref{fig:JCM_typical_decay} can be recognized as the semi-classical Rabi oscillations of the effective Rabi frequency $\frac{\Omega\mathcal{G}}{2\Delta_e}|\alpha|$. Moreover, the re-emission from the ensemble's collective excitation into the cavity mode is practically a process of superradiance \cite{PhysRevLett.102.143601}. Therefore, the Rabi oscillation can alternatively be viewed as the absorption and re-emission of the photonic state in the cavity mode by the many intracavity emitters.

The discussions so far have hinted that such a system allows not only the study of JCM in an interesting parameter sector but also potential applications in spin squeezing and superradiant lasing. The role of Rydberg blockade is essential since it guarantees that the entire atomic ensemble behaves a lot more like a single emitter. This ensures that the quantum Rabi oscillation is taking place even when it is driven by an optical coherent state which can be regarded as the classical field. On the contrary, a classical optical pulse driving a medium of uncorrelated emitters is hardly capable of yielding this demonstrated behavior. With the contents discussed in this section, it can be observed that the proposed stimulated Raman approach with two-photon transition has several unique characteristics, when compared with the approaches of intracavity Rydberg EIT \cite{Hao2015srep, PhysRevA.93.040303, PhysRevA.94.053830} or the single atom coupled to a high-finesse cavity \cite{PhysRevLett.92.127902, Reiserer13177}.

\section{Atom-photon quantum gate}
\label{sec:gate}

\subsection{Controlled-PHSAE gate via Rydberg blockade}
\label{sec:phase_shifts}

The isomorphism to JCM discussed in Section \ref{sec:framework} already hints that the system of intracavity Rydberg-blocked atomic ensemble may have potential applications for atom-photon entanglement in quantum optics, where the cavity resonance frequency is detuned by a significant amount from the atomic resonance frequency. Inspired by the strong-coupling regime of the atom-photon interaction discussed in Section \ref{sec:JCM}, the analogy with the typical cQED scenario leads to a straightforward observation that a controlled-PHASE gate can be accordingly constructed on the same platform of this system.

\begin{figure}[t]
\centering
\includegraphics[width=\linewidth]{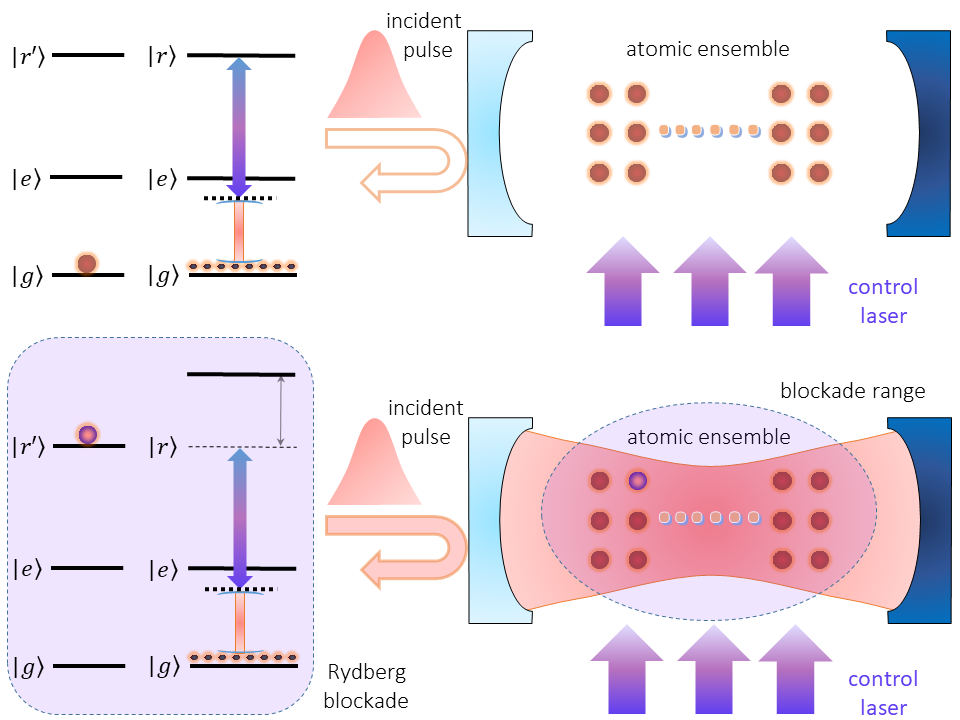}
\caption{An outline for the basic principles of the atom-photon gate with intracavity Rydberg-blocked atomic ensemble, where the qubit state on the atom side is abstracted into the internal states of a single atom within the ensemble. The optical cavity is supposed to be single-sided, where the one end mirror is perfectly reflecting. The frequency of the incident single-photon pulse is resonant with the optical cavity, but not necessarily so with the atomic transition $|g\rangle \to |e\rangle$. This graph is for the particular case where the matter qubit is instantiated in the form of a single atom among the atomic ensemble. In such a configuration, strong Rydberg blockade is presumed to take place between state $|r'\rangle$ of the qubit atom and state $|r\rangle$ for the rest atoms of the ensemble, as a consequence of the F\"{o}rster resonance structure in \eqref{H_CZ_1}. Nevertheless, the single qubit atom does not have to be same species as the other atoms in the ensemble \cite{PhysRevA.94.053830, PhysRevLett.119.160502}.}
\label{fig:basic_gate}
\end{figure}

The basic idea of the mechanism is illustrated in Fig. \ref{fig:basic_gate}, the purpose of which is to establish a controlled-Z quantum gate between photonic qubit and matter qubit. The instantiation of the matter qubit may be in the form of a distinguished single atom in the ensemble \cite{PhysRevA.92.022336, PhysRevLett.119.160502, J.Phys.B.49.202001} or a spin wave embedded in the entire ensemble \cite{Choi06670, PhysRevLett.95.133601, Dudin887Science, PhysRevLett.115.093601}, where in both cases the qubit register states can be chosen as the hyperfine states of the ground level for typical alkali atoms and the gate protocol can be chosen to include the excitation from ground state to Rydberg state as the very original proposal \cite{PhysRevLett.87.037901, RevModPhys.82.2313}. As shown in Fig. \ref{fig:basic_gate}, the single-photon pulse incidence is ultimately reflected from the cavity, where it gains a phase shift depending on the matter qubit's state during the process. 
The Rabi frequency and detuning of the control laser is supposed to ensure that the two-photon transition together with the intracavity optical field is on resonance for the atoms, whose details will be discussed quantitatively later. For simplicity, let's assume the two register states of the matter qubit is $|g\rangle$ and $|r'\rangle$. When the matter qubit is sitting at the state $|g\rangle$, it does not exert any substantial impact on the other atoms, and the two-photon transition from $|g\rangle$ to $|r\rangle$ exactly holds. Therefore in this case the cavity resonance frequency is effectively shifted and henceforth the incident single-photon pulse can not enter the cavity \cite{PhysRevLett.92.127902}. When the matter qubit is sitting at the state $|r'\rangle$, it influences the rest of atoms via Rydberg blockade such that the two-photon transition for the cavity field and control laser is out of resonance. Therefore the incident single-photon pulse enters the cavity freely before it gets reflected eventually. Briefly speaking, upon reflection the single-photon pulse gains a conditional phase shift, 0 or $\pi$, depending on whether the matter qubit state is $|g\rangle$ or $|r'\rangle$. 

Here we begin a quantitative analysis for this gate protocol. The interaction Hamiltonian for the atom-photon interaction of this intracavity atomic ensemble system including the Rydberg blockade effects is:
\begin{align}
\label{H_CZ_1}
H_{p1} = 
&\sum_{n=1}^{N}(-\hbar\frac{\Omega}{2}|r_n\rangle\langle e_n| - i\hbar \mathcal{G}_n|e_n\rangle\langle g_n| \hat{b}) + \text{H.c.} \nonumber\\ % couplings
&-\hbar\Delta_e\sum_{n=1}^{N} |e_n\rangle\langle e_n| 
-\hbar\Delta_r\sum_{n=1}^{N} |r_n\rangle\langle r_n| 
\nonumber\\ % effective detunings
&+ \hbar\sum_{n=1}^{N} V_n|r_n\rangle\langle p_n| \otimes |r'\rangle\langle p'| + \text{H.c.} 
\nonumber\\
&+ \hbar\delta_p \sum_{n=1}^{N} |p_n\rangle\langle p_n| \otimes |p'\rangle\langle p'|,
\end{align}
where $|r'\rangle, |p'\rangle$ are Rydberg states of the qubit atom, $|r\rangle, |p\rangle$ are Rydberg states of the other atoms and those two pairs of states form the F\"{o}rster resonance $|rr'\rangle \leftrightarrow |pp'\rangle$. $\delta_p$ is the small F\"{o}rster energy penalty term from the Rydberg product states $|p\rangle|p'\rangle$.

From a realistic point of view, the values of $\Delta_e$ and $\mathcal{G}_n$ are fixed to begin with, as the cavity hardware is already chosen and incoming sing-photon pulse's frequency need to be resonant with the cavity. On the other hand, flexibility is granted for the control laser in terms of $\Omega$ and $\Delta_r$. For example, under the specific condition of the typical Autler-Townes effect, it is fair to set $\Delta_e = \Delta_r = \Delta$.

The state vector for the complete system is similar to the form described by \eqref{JCM_state_vector_1}. Note that the difference with respect to \eqref{JCM_state_vector_1} is that now the state may definitely contain one intracavity Rydberg excitation in the qubit atom.
\begin{align}
\label{gate_state_vector_1}
|\Psi(t)\rangle  = & \int d\omega \, \phi_s (\omega, t) \hat{a}^\dagger_s(\omega) |g^N, r', \O_b, \text{\O}_a\rangle  
\nonumber\\
&+ C_b(t) \hat{b}^\dagger |g^N, r', \text{\O}_b, \text{\O}_a\rangle
\nonumber\\
&+ \sum_{m=1}^{N} C_{em}(t) |g^{N-1}, e_m, r', \text{\O}_b, \text{\O}_a\rangle \nonumber\\
&+ \sum_{m=1}^{N} C_{rm}(t) |g^{N-1}r_m, r', \text{\O}_b, \text{\O}_a\rangle
\nonumber\\
&+ \sum_{m=1}^{N} C_{pm}(t) |g^{N-1}p_m, p', \text{\O}_b, \text{\O}_a\rangle.
\end{align}

Based upon \eqref{H_CZ_1} and \eqref{gate_state_vector_1}, we proceed to compute the conditional phase shifts according to the quantum input-output theory together with quantum jump approach. The control laser parameters are assumed to be time-independent, and then the formal calculation is to be carried out on the Fourier domain. From this point on, $\delta$ is used to label the frequency components of the incident single-photon pulse, where $\delta=0$ denotes a component that's completely resonant with the cavity. 

We make a few definitions here: the ac Stark shift induced by the atom-photon interaction on the ground level $\Delta_{ac}$, the dressing factor to the atom-photon coupling strength $\eta$, the effective relative frequency shift for coupling to the dressed states $\Delta_{dr}$, and the effective Rydberg blockade shift $B_m$ where the subscript $m$ is the index for the atom. 
\begin{align}
\label{gate_shifts_def}
\Delta_{ac} = - \frac{\mathcal{G}^2}{\Delta_e +\delta + i\frac{\Gamma_e}{2}},\,
\eta = \frac{1}{4}\frac{|\Omega|^2}{(i\Delta_e - \frac{\Gamma_e}{2} + i\delta)^2},\, \nonumber\\
\Delta_{dr} = -\Delta_r 
-\frac{i}{4} \frac{|\Omega|^2}{\frac{\Gamma_e}{2} - i\Delta_e -i\delta}, \,
B_m = \frac{|V_m|^2}{\frac{\Gamma_p}{2}+i\delta_p-i\delta}.
\end{align}

With the above preparations, the conditional phase shifts can be readily expressed. When the qubit atom is sitting at the state $|r'\rangle$ the cavity, the reflection coefficient is:
\begin{equation}
\label{gate_R1}
% \big \Big \bigg \Bigg
R_\odot(\delta) = 1-
\kappa \Big\{ \frac{\kappa}{2} - i\delta - i\Delta_{ac}
- \eta \sum_{m=1}^{N} 
\frac{|\mathcal{G}_m|^2}{\frac{\Gamma_r}{2} - i\delta + i\Delta_{dr} + B_m}
\Big\}^{-1}.
\end{equation}
On the other hand, when there is no stored excitation in the qubit atom, the reflection coefficient reads:
\begin{equation}
\label{gate_R0}
% \big \Big \bigg \Bigg
R_\oslash(\delta) = 1-
\kappa \Big\{ \frac{\kappa}{2} - i\delta - i\Delta_{ac}
- \frac{ \eta \mathcal{G}^2 }{\frac{\Gamma_r}{2} - i\delta + i\Delta_{dr}}
\Big\}^{-1}.
\end{equation}

In the limit that $\mathcal{G}^2/(\delta\kappa) \to \infty, (\mathcal{B}\kappa)/\mathcal{G}^2 \to \infty, (\Delta_e\kappa)/\mathcal{G}^2 \to \infty$ where $\mathcal{B}$ conceptually denotes the effective Rydberg blockade shift, the conditional phase shifts obey $R_\odot \to -1, R_\oslash \to 1$ and the performance of this controlled-Z gate is ideal. Recent advances in utilizing the dark state to enhance the fidelity of atom-atom controlled-PHASE gate \cite{PhysRevA.96.042306} are also expected to offer future improvements of the atom-photon gate under discussion here (see also a recent advance in utilizing dark state technology to improve atom-photon gate: \emph{arXiv: 1801.07241}).

The above analysis together with \eqref{gate_R1} and \eqref{gate_R0} is also essentially applicable to the case where the matter qubit is realized via a spin wave embedded in the entire atomic ensemble. The encoding (write) and retrieval (read) of a single excitation in the form of spin wave within the atom ensemble inside the optical cavity is robust and efficient, where recent experimental progress has confirmed that a single excitation of spin wave in intracavity atomic ensemble can be efficiently prepared and retrieved \cite{PhysRevLett.95.133601, PhysRevLett.112.033601}, allowing single-photon pulse to be coupled out from the atom-cavity system with near unity efficiency. As long as state $|r'\rangle$ differs from state $|r\rangle$, which is common in the state choices in implementing Rydberg blockade, the operational laser frequency for controlling the spin wave states and the laser frequency for controlling the two-photon transition of the gate protocol will usually be more than a few hundred GHz apart, and therefore the crosstalk can be made minimal. Moreover, this gate protocol can work as an atom-photon quantum gate for single-photon pulses endowed with polarization encodings, as well as the time-frequency encodings such as frequency-bin encoding \cite{PhysRevA.97.023839}.

This process is natural in the sense that its physic interpretation is closely related to the original proposal of atom-photon Controlled-Z gate \cite{PhysRevLett.92.127902, RevModPhys.87.1379}. In the mean time it also contains more practical flexibility compared with the previously known schemes, in particular, there exists a relatively wide tuning range for the cavity resonance frequency $\Delta_e$, and therefore this new gate protocol can even be used as a photon-photon gate to entangle single-photon pulses with different frequencies up to a few hundred MHz. This tuning ability and the choice of sizable $\Omega, \Delta_e$ is also a distinguished feature where we focus on the two-photon transition and Autler-Townes effect, when compared with the recent advances reported in \emph{arXiv: 1801.07241} which focuses on Rydberg dressing and dark states of Rydberg EIT system.

\subsection{Estimation of gate performance}
\label{sec:performance}

\begin{figure}[ht]
\centering
\fbox{\includegraphics[width=\linewidth]{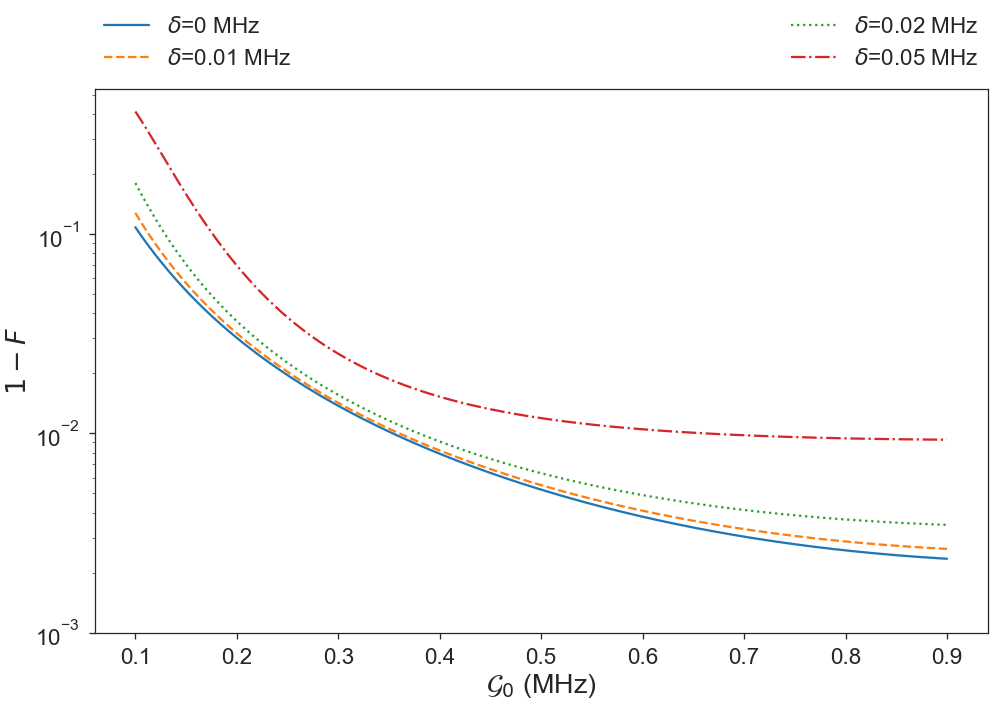}}
\caption{Numerical simulation of the gate's fidelities with respect to different the single-atom single-photon coupling strength $\mathcal{G}_0$. Particular parameters for this simulation includes: $\Omega=2\pi\times100\text{ MHz}, \Delta_e=2\pi \times 1000 \text{ MHz},\Gamma_e = 2\pi \times 1 \text{ MHz}, \Gamma_r = \Gamma_p = 2\pi \times 0.01 \text{ MHz}, \kappa = 2\pi \times 1 \text{ MHz}$.}
\label{fig:gate_fidelity_1dscan}
\end{figure}

In this sub-section, the estimated performance and related numerical simulations of the atom-photon Controlled-Z gate are provided. Since we have already checked the conditional phase shifts of the reflection coefficient, we may then calculate the Choi-Jamiolkowski fidelity \cite{JAMIOLKOWSKI1972275, CHOI1975285, PhysRevA.71.062310, Hao2015srep, PhysRevA.93.040303, PhysRevA.94.053830} of the proposed gate operation. According to the arguments made in Refs. \cite{PhysRevA.93.040303, PhysRevA.94.053830}, the average fidelity for the atom-photon quantum gate is:
\begin{equation}
\label{gate_fidelity_def}
F_z = \frac{1}{16}|2 + R_\oslash - R_\odot|^2. 
\end{equation}

Next, we are going to compute $F_z$ via numerical simulations for typical parameter settings. The intracavity ensemble is set as a 3D atomic array of 10 by 10 by 10 with site spacing being 0.37 $\mu$m. The single-atom qubit is placed at 1.5 sites away from the top layer center of the array. The Rydberg-Rydberg interaction parameters are taken from $^{87}$Rb atoms, where $|r\rangle$ is regarded as 81S and $|r'\rangle$ is regarded as 84S. For larger principal quantum number, it is possible to get even stronger F\"{o}rster resonance. The single-atom single-photon coupling is assumed to be the same as $\mathcal{G}_0$ for every atom in the ensemble. The angular dependence of the Rydberg-Rydberg interaction is also taken into consideration according to Ref. \cite{PhysRevA.92.042710}. If the qubit state is realized via a spin wave embedded in the entire atomic ensemble, the gate fidelity results will be similar, whose details are omitted here for simplicity.

\begin{figure}[htbp]
\centering
\fbox{\includegraphics[width=\linewidth]{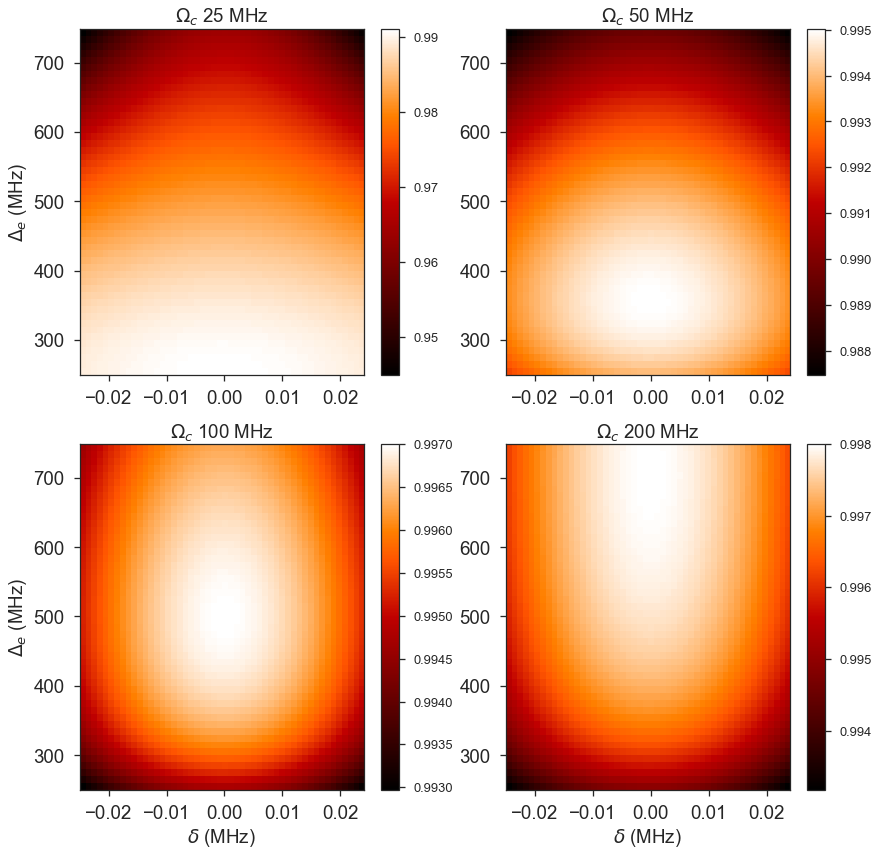}}
\caption{Numerical simulation of the gate's fidelities to demonstrate its tuning range in $\Delta_e$, with respect to different Rabi frequency settings of the control laser. Particular parameters for this simulation includes: $\mathcal{G}_0 = 2\pi \times 0.5 \text{ MHz}, \Gamma_r = \Gamma_p = 2\pi \times 0.01 \text{ MHz}, \kappa = 2\pi \times 1 \text{ MHz}$.}
\label{fig:gate_fidelity_2dscan}
\end{figure}

Sample results of the numerical simulations are presented in Fig. \ref{fig:gate_fidelity_1dscan} and Fig. \ref{fig:gate_fidelity_2dscan}. As already been revealed by previous theoretical deductions, we observe that the implementation of this gate protocol does not require ultra-high finesse from the optical cavity to get strong effective atom-photon coupling. Meanwhile, it clearly offers a rather versatile parameter range. More specifically, it is able to serve as an interface where the photon frequency is endowed with a reasonable frequency dynamic range.

A relatively straightforward way to comprehend the working principle of this gate protocol is the Autler-Townes effect. In the strict sense, this is referring to the situation that $\Delta_e$ is the same as $\Delta_r$ and $\Omega = 2\Delta_e$ while discounting the ac Stark shifts. The two dressed states of the ensemble atoms are $(|e\rangle + |r\rangle)/\sqrt{2}$ and $(|e\rangle - |r\rangle)/\sqrt{2}$, where only one of them is hitting the resonance to prevent the single-photon pulse from entering the cavity. Otherwise, if the Rydberg blocked is turned on, this resonance with one of the two dressed states is broken and the single-photon pulse may enter the cavity. If the system is liberated from the very strict definition of Autler-Townes effect, it can be extended from this special case to the general concept of two-photon stimulated Raman transition where the magnitude of $\Omega$ is a sizable fraction of $|\Delta_e|$, not necessarily 2 or 1. This is tied to the relative strength of the state $|e\rangle$ in the dressed state which is actually resonantly coupled to the ground state.

With such a configuration to implement the atom-photon controlled-PHASE gate, several potential upgrades can be considered towards reaching an even higher fidelity. They include the reduction of the effective intermediate state decay $\Gamma_e$, choosing Rydberg states with longer lifetimes, Stark tuning to obtain stronger F\"{o}rster resonance, and increasing the number of atoms in the ensemble. Another point calling for caution is the sign of the Rydberg blockade shift. If the magnitude of the Rydberg blockade shift is so large such that the intermediate detuning $\Delta_e$ on longer dominates, its sign shall be chosen as pulling the system towards a direction away from both the two dressed states' resonances.

\section{Conclusion and outlook}
\label{sec:conclusion}

In summary, we have offered the analysis for the interaction between intracavity Rydberg blocked atomic ensemble and cavity optical fields, with the emphasis put on the quantum optical properties of the dynamics. The fundamental relation with JCM is explored, where the atoms participate the atom-photon interaction collectively with cavity field due to Rydberg blockade. Moreover, an atom-photon contolled-PHASE quantum gate is constructed on top of the insights gained through the study of JCM for this hybrid system, whose performance is also investigated. The exact number of atoms inside the ensemble does not play an essential role in the dynamics of the system as long as the total number is adequate to enhance the collective atom-photon coupling, while the functionality of the system is robust against atom loss \cite{PhysRevA.62.062314}. The behavior of this hybrid system clearly shows signatures from the well-known Autler-Townes effect and two-photon stimulated Raman transition. More specifically, the cavity optical field couples the atom ground state to a dressed state made from an intermediate state $|e\rangle$ and a Rydberg state $|r\rangle$, which is not the same case as a typical cQED scenario where a single two-level atoms is coupled with the cavity optical field. This feature enables the exploration of parameter regions not so easy to access previously, from the cQED point of view.

With a high fidelity controlled-phase gate between the photonic polarization qubit and atomic spin qubit, the potential exists for designing a quantum repeater based upon our work, which will extend photonic entanglement to very long distances. One essential problem for the atom-photon gate in the real world is its fidelity under operational conditions where technical noises and systematic imperfections are always present. Therefore, we want to emphasize that the long standing goal of efficiently realizing high fidelity atom-photon and photon gates remains as a difficult challenge, where enormous theoretical and experimental efforts are devoted to the research for this topic. We hope that our work offers help to this pursuit.

\section*{Funding Information}
National Key R\&D Program of China (under contract Grant No. 2016YFA0301504).

\section*{Acknowledgments}
The authors gratefully acknowledge the funding support from the National Key R\&D Program of China (Grant No. 2016YFA0301504) and the basic research program fund at ICQI of National University of Defense Technology. The authors also acknowledge the hospitality of Key Laboratory of Quantum Optics and Center of Cold Atom Physics, Shanghai Institute of Optics and Fine Mechanics. 
The authors gratefully thank Professor Mark Saffman whose help offers enormous momentum to this work. The authors also thank Professor Peng Xu and Professor Xiaodong He for enlightening discussions.

\bibliographystyle{apsrev4-1}

\renewcommand{\baselinestretch}{1}
\normalsize

%\clearpage%
%\phantomsection%
%\addcontentsline{toc}{chapter}{\numberline{}{Bibliography}}%
\bibliography{interface_ref}
\end{document}